\documentclass[12pt,preprint]{aastex}

\begin{document}

\title{THE ABUNDANCE OF FLUORINE IN NORMAL G AND K STARS 
OF THE GALACTIC THIN DISK}

\author{C. A. Pilachowski\altaffilmark{1} \& Cameron Pace\altaffilmark{1}}
\affil{Astronomy Department, Indiana University Bloomington,
    Swain West 319, 727 East Third Street, Bloomington, IN 47405-7105, 
USA; cpilacho@indiana.edu, cjpace@indiana.edu}

\altaffiltext{1}{Visiting Astronomer, Kitt Peak National Observatory.  
KPNO is operated by AURA, Inc.\ under contract to the National 
Science Foundation.}

\begin{abstract}
The abundance of fluorine 
is determined from the (2-0) R9 2.3358 $\mu$m 
feature of the molecule HF for several dozen normal G and K stars in the 
Galactic thin disk from spectra obtained with the Phoenix IR spectrometer 
on the 2.1-m telescope at Kitt Peak. 
The abundances are analyzed in the context of Galactic chemical evolution
to explore the contributions of supernovae and asymptotic giant branch
(AGB) stars to the abundance of fluorine in the thin disk.  
The average abundance of fluorine in the thin disk is found to be
[F/Fe]~=~+0.23~$\pm$~0.03, and the [F/Fe] ratio is flat or declines slowly with 
metallicity in the range from --0.6~$<$~[Fe/H]~$<$~+0.3, within the limits
of our estimated uncertainty.  The measured abundance of fluorine and
lack of variation with metallicity in Galactic thin disk stars suggest
neutrino spallation in Type II supernovae contributes significantly to 
the Galactic fluorine abundance, although
contributions from AGB stars may also be important.

\end{abstract}

\keywords{stars: abundances; stars: late type; Galaxy: abundances; 
Galaxy: disk}

{\it Facility:} \facility{KPNO: 2.1m (Phoenix)}

\newpage

\section{Introduction}

The abundances of a variety of chemical elements contribute to our
understanding of the chemical evolution of the Milky Way Galaxy,
including Li and CNO (Spite et al., 2005), the alpha-process elements 
and transition peak metals (Tinsley 1980), 
and the n-capture species (Sneden et al. 2003).
The origin of the abundant, even-Z, light elements through Ca (and perhaps Ti)
are reasonably well-understood through helium, carbon, neon, and oxygen
burning, the silicon quasi-equilibrium process, and subsequent
explosive nucleosynthesis in Type II supernovae (SNe II; Nomoto et al. 2013).
The origin of the less abundant, odd-Z elements Na through Cl, particularly 
fluorine, phosphorus, and chlorine, is less well understood, yet these 
elements provide an opportunity to constrain secondary nucleosynthesis
processes operating on more abundant species in SNe II supernovae. 
The odd-Z element abundances can also be modified through
proton-capture reactions in low- and intermediate-mass asympotic giant branch 
(AGB) stars (see, e.g. Cristallo et al. 2009; Ventura \& D'Antona 2008), 
as seen, for example, in the unusual
chemical evolution processes operating in globular clusters
(e.g. Ventura et al. 2001; Gratton et al. 2012).

Briefly, the light, odd-Z element fluorine can be produced in a variety of 
astrophysical environments; for a full discussion of sources of fluorine, 
see recent papers by Kobayashi et al. (2011), Recio-Blanco et al. (2012), and
J{\"o}nsson et al. (2014a).
Potential sources in the Galaxy include SNe II, AGB stars, and possibly 
Wolf-Rayet stars
(production of fluorine in Type Ia supernovae is thought to be small).
Production in SNe II, however, is not sufficient to account for the abundance of
fluorine in the Galaxy without the additional production due to the
$\nu$-process involving inelastic scattering of $\mu$ and $\tau$ neutrinos 
off $^{20}$Ne (Woosley et al. 1990, Woosley et al. 2002,
Kobayashi et al. 2011). 

In AGB stars, fluorine is produced in core and shell helium-burning, but 
is destroyed by proton capture at the base of the convective envelope for stars
with initial masses of 4-7 solar masses, and by alpha captures at temperatures
above 2.5 x 10$^{8}$ K (see Lugaro et al. 2004 \& 2012, Karakas et al. 2008, 
and Gallino et al. 2010).
Thus, significant amounts of fluorine can be contributed only 
by AGB stars in the mass range 
2-4 solar masses, so that fluorine from AGB stars should only be found in
stars with metallicities greater than [Fe/H] = --1.5 (Kobayashi et al. 2011).
We note, however, that more masssive AGB stars 
(see e.g. Garc\'{i}a-Hern\'{a}ndez et al. 2006, 2007)   
could overproduce fluorine to a smaller extent once hot bottom burning 
ceases (Karakas et al. 2010, Lugaro et al. 2012, D'Orazi et al. 2013).  
The enrichment of fluorine is less in more massive AGB stars but they
expel much more material to the interstellar medium than do lower mass AGB 
stars.

Fluorine abundances in AGB stars have been reported in numerous studies,
beginning with Jorissen et al. (1992), which has since been partly revised to 
lower values by Abia et al. (2009, 2010).  
More recently, fluorine has been found in stars contaminated with 
products of AGB nucleosynthesis via mass transfer.  
Schuler et al. (2007) and Lucatello et al. (2011), for example, found enhanced
fluorine in some carbon-enhanced metal-poor (CEMP) stars, which they
attribute to contamination by former AGB companions.
In addition, some authors have reported high fluorine abundances in
second generation stars in globular clusters
(see, for example, de Laverny \& Recio-Blanco 2013, D'Orazi et al. 2013,
and model predictions by Ventura \& D'Antona, 2008).
However, Kobayashi et al. (2011) concluded that AGB production of fluorine 
is not sufficient to account for the observed abundance of fluorine at the 
solar metallicity. 
Also of interest is the observation of fluorine enhancements in
R Coronae Borealis and extreme helium stars (Pandey 2006, Pandey et al. 2008,
and Jeffery et al. 2011).

Production in Wolf-Rayet (WR) stars during the early He-burning phase is also
possible (e.g. Meynet \& Arnould 1993, 2000), but the fluorine must be 
returned to the interstellar medium though  stellar winds before temperatures
in the He-burning zone exceed 2.5 $x$ 10$^{8}$ K where fluorine is destroyed.
Palacios et al. (2005) reexamined the production of fluorine by WR stars using 
newer yields and models including rotation, but noted that major uncertainties
in yields remain.
J{\"o}nsson et al. (2014a) suggested that contributions from WR stars may  
be required to account for trends in both [F/Fe] and [O/Fe] in Galactic
bulge giants.
   
Observations of the Galactic abundance of fluorine require the determination
of the fluorine abundance in less-evolved field stars not contaminated by AGB
nucleosynthesis.
Jorissen et al. (1992) included a handful of normal giants, and their sample
was expanded by Cunha \& Smith (2005) and by Recio-Blanco et al. (2012).
More recently, J{\"o}nsson et al. (2014b) published analyses of the fluorine
abundance in six nearby, cool giants (plus the thick disk giant Arcturus).
In this paper, we report the abundance of fluorine in several dozen field
giants and dwarfs residing in the Galactic thin disk, in order to determine
the Galactic abundance of fluorine as a function of metallicity and to 
constrain the nucleosynthetic sources of fluorine contributing to chemical
enrichment in the Milky Way.
In Section 2, we describe the selection of targets, observational material, and
analysis, and compare our results to the literature.  
In Section 3, we discuss our results in the context of the Galactic
thin disk and chemical evolution models, and finally, in Section 4, we summarize
our conclusions.

\section{Observations and Analysis}

\subsection{Target Selection}

Two criteria were used to select stars to be included in this study.
First, stars must have available atmospheric parameters, including temperature,
surface gravity, and [Fe/H] from the literature, specifically from the
Pastel Catalogue of Stellar Parameters (Soubiran et al. 2010).  Atmospheric 
parameters are needed because the limited spectral region available with our 
Phoenix observations is not sufficient to determine spectroscopic parameters
independently. Second, stars must be probable members of the 
Galactic thin disk.

The probablility of stars to belong to the thin disk was estimated following 
the methodology of Johnson \& Soderblom (1987), as described by
Ram\'{i}rez et al. (2013).  For most stars, the U, V, and W space
velocities were obtained from Casagrande et al. (2011); for stars not included
in this catalog, we used available proper motion, parallax, and radial
velocity data from SIMBAD to calculate the space motions, using the online
tool provided by David Rogriguez\footnote
{http://www.das.uchile.cl/~drodrigu/UVWCalc.html; August 2013}.  
From the U, V, and W space velocities, the probability of membership in the thin disk,
thick disk, and halo populations can be calculated; stars with probabilities
of belonging to the thin disk of greater than 50\% were then included in our 
analysis.  Most stars in our sample have probabilities greater than 90\%
of belonging to the thin disk population, although six stars have membership
probabilities in the range 0.75$<$P$<$0.9 (HD 5268, HD 17660, HD 32147, 
HD 37984, HD 218031, and HD 222107) and two stars have less certain 
membership (HD 43039 at P=0.56 and HD 39715 at P=0.67). 

\subsection{Observations}

Spectra of the 2.336 $\mu$m micron region containing a reasonably unblended
feature of the molecule HF were obtained with the Phoenix spectrometer
(Hinkle et al. 1998)
on the 2.1-m telescope of the Kitt Peak National Observatory in 2012
November and December.  
This spectral region is dominated by strong lines of CO, with relatively 
few atomic lines.  The spectrograph was configured with a 4-pixel (107 $\mu$m)
slit corresponding to 0.7 arcsec on the sky.  With the 4308 filter to isolate
grating order 32, we obtained a spectral resolving power of 25,000 and
spectral coverage from 2.3285-2.3390 $\mu$m.  Observations were obtained in 
pairs at two slit positions to facilitate dark current and sky subtraction.
Typically four observations were obtained for each star at two different
slit positions to remove thermal emission from the sky.
To minimize telluric absorption, stars were observed close to the meridian,
with zenith distances typically less than 25 degrees (airmass $<$ 1.1).
Telluric lines were removed using the IRAF\footnote{IRAF is distributed
by the National Optical Astronomy Observatory, which are operated by the
Association of Universities for Research in Astronomy, Inc., under
cooperative agreement with the National Science Foundation.} task telluric, 
which shifts and scales the telluric line spectrum to minimize residual 
effects of telluric lines.  
Telluric line division was specifically optimized for the spectral 
region near the HF feature.  
Each star's radial velocity shifts its HF
feature relative to the telluric spectrum, so that each star is affected
differently.  

\subsection{Analysis}                   
The abundance of fluorine was determined using both spectrum synthesis
and equivalent width analysis of the 2.3358 $\mu$m feature. 
The equivalent width method provides a more robust upper limit in cases where
the HF feature is not detected, or detection is questionable.
The analysis utilized the LTE spectrum synthesis code Moog (Sneden 1973, 2010
version) and model atmospheres interpolated in
the MARCS\footnote{The interpolation of models utilized code provided by
Masseron (2006, http://marcs.astro.uu.se/software.php).} grid 
(Gustafsson et al. 2008). 

Following D'Orazi et al. (2013), an excitation potential of $\chi$=0.227~eV 
was adopted for the 2.3358 $\mu$m feature of HF from the HITRAN molecular
line database (Rothman et al. 2013).
As discussed by Nault \& Pilachowski (2013), this value differs from the value 
of 0.49 eV previously used in the literature (including in the original solar
abundance determination by Hall \& Noyes 1969), and results in a lower
abundance of fluorine by typically 0.36 dex\footnote{We use the standard 
spectroscopic notation where 
[A/B]$\equiv$log(N$_{\rm A}$/N$_{\rm B}$)$_{\rm star}$
~--log(N$_{\rm A}$/N$_{\rm B}$)$_{\sun}$ and
log $\epsilon$(A)$\equiv$log(N$_{\rm A}$/N$_{\rm H}$)+12.0 
for elements A and B.}. 
The oscillator strength log~gf~=~--3.971 was adopted from Lucatello et al.
(2011);  this value is close to the value typically used in the literature
(see, for example, Jorissen et al. 1992 and Recio-Blanco et al. 2012).
The dissociation energy used by Moog is 5.8698 eV.
J{\"o}nsson et al. (2014a) carefully examined the calculation of the partition
functions for HF, and J{\"o}nsson et al (2014b) concluded that the partition
functions used with Moog are in agreement with their calculations.

For spectrum synthesis of the neighboring CO (2--0) and (3--1) 
vibration-rotation lines, we adopted wavelengths, excitation potentials, 
and gf-values from Goorvitch (1994).  
A handful of atomic lines are present in the spectrum,
and we adopted line parameters from the Vienna Atomic Line Database (VALD)
(Kupka et al 2000, and references therein) for our spectrum synthesis. 

Since our spectral region is limited, with few atomic lines, we were unable
to derive reliable stellar model atmosphere parameters from our spectra.
Instead, we adopted temperatures, surface gravities (log g), values of 
the microturbulence parameter ($\xi$), and metallicities from the literature,
selecting sources whereever possible from large compendia.  
The adopted atmospheric parameters are included in Table 2 for all stars 
in our sample.  
The most frequently used sources for atmospheric parameters
are the large samples studied by McWilliam (1990) and Prugniel et al. (2011), 
although several additional sources were needed to identify parameters 
for all stars in our sample.  
All sources are listed in the references for Table 2. 

A synthetic spectrum of the HF region was computed for each star using the 
adopted model atmosphere parameters.  Initial abundances for CNO were estimated 
based on the stellar metallicity and spectral type, and then adjusted to 
match the observed CO line strengths.  
Since the abundances of neither C nor O are available in the literature for
most of our program stars, we are unable to determine abundances of these
species from the CO lines alone.  Once the CO spectrum was fit,
the abundance of fluorine was adjusted to match the observed HF line profile.
Sample observed and synthetic spectra are shown in Fig. 1.

The nearby (2-0) R25 line of C$^{12}$O$^{17}$ is present in some spectra, 
and was included in the synthesis calculation to account for possible blending
with the HF feature.  The O$^{16}$/O$^{17}$ ratio varies from a few hundred in 
giant stars
to a few thousand in dwarf stars (Clayton 2003), and the presence of possible
contamination from C$^{12}$O$^{17}$ can be estimated from other stronger, 
unblended features in the observed spectral range.

Equivalent widths of the relatively isolated 2.3358 $\mu$m feature of HF
were also measured from the observed 
spectra using the \emph{splot} task in IRAF with a Gaussian fit to the line.
If present, the  C$^{12}$O$^{17}$ feature was also fit with a Gaussian to
eliminate its contribution to the HF equivalent width.

The \emph{abfind} driver from Moog was used to obtain the abundances of 
fluorine 
from the measured equivalent widths using the same 
model atmosphere as for the syntheses.  
Values of the equivalent width of HF range from lower limits of 
$<$10~m\AA\  to 250~m\AA\  (log~W/$\lambda$=--4.97).
The measured equivalent widths or equivalent width upper limits are
included in Table 2.

The two methods of analysis provide similar results for fluorine, typically
within 0.2~dex in log~$\epsilon$(F), with an average difference of 0.09~dex.  
Sources of uncertainty from the spectrum synthesis include both the
continuum level and the smoothing factor to match the instrumental profile.
Uncertainties in equivalent widths are dominated by the continuum height, as 
well as noise for cases with weak HF lines or lower S/N ratio.
Both methods are affected similarly by uncertainties in the atmospheric
parameters and atmospheric modeling. 
The two results were averaged in most cases.
When the two results were discrepant, an overly optimistic fit of the 
synthetic spectrum to noise was generally the cause;  in these cases,
the equivalent width provides a more realistic upper limit. 

The final adopted values of the abundance of fluorine are 
included in Table 2, and are plotted versus effective temperature in Fig.~2.
The abundance of fluorine is flat across a range of stellar
temperatures, except at the warm end above 4600~K, where the HF
feature becomes too weak and most measurements are upper limits unless the 
fluorine abundance is high. 
The HF feature disappears in stars hotter than about 4700~K due to 
molecular dissociation, so our effective sample is limited to stars with 
temperature below this limit.  Our sample may contain an imcompleteness bias
for stars with 4500~K $<$ $T_{eff}$ $<$ 4700, since detection of HF depends on 
both the S/N ratio of the spectrum and the metallicity of the star.
The upper limit for HD 220009 at $T_{eff}$~=~4314K results from the low
metallicity of the star at [Fe/H]~=~--0.7.
The apparent rise in log~$\epsilon$(F) with temperature is likely due to
observational limitations rather than a systematic temperature error.

The M0Iab supergiant HD 216946 (HR 8726) is labelled in Fig. 2.
The star exhibits an anomalously high fluorine abundance, and may
be affected by in situ proton capture nucleosynthesis.
The star will not be considered further here.
                                                                                                                             
Li et al. (2013) have considered deviations from radiative equilibrium in
the formation of the HF R9 line, specifically due to 3D effects in
the stellar atmospheres.  
The investigations of Li et al. apply in the low metallicity regime near 
[M/H]~=~--2.0. They found that 3D abundance corrections for HF are small 
($<$ 0.03 dex) at low gravity and temperature but increase with gravity 
and temperature to values near 0.42 dex at 5024 K and log~g~=~2.5.
Asplund (2005) notes that the corrections to apply to 1D models
are more significant at lower metallicity, although such effects are
generally more significant for molecular features than for atomic features.
The stars considered here are all near-Solar in abundance, so corrections
should be less for these stars than for metallicities near [Fe/H]~=~--2,
and the absence of a dependence of derived abundance on
temperature in Figure 2 gives some confidence that 3D effects do not increase
significantly with temperature for stars below 4500 K.
Restricting our sample to stars with effective temperatures below 4500 K,
we also note that the average fluorine abundance for thin disk giants is 
[F/Fe]~=~+0.23~$\pm$~0.03 (standard error of the mean (SEM)),
while the average abundance in the dwarfs is [F/Fe]~=~+0.19~$\pm$~0.12
(SEM), likely the same within our estimated uncertainties.

Observational uncertainties in the fluorine abundance can be estimated from 
discernable differences in the synthetic spectrum fits ($\pm$0.05 dex).
Additional uncertainties, both random and systematic, arise from the
selection of atmospheric parameters.
Among the various atmospheric parameters, the abundance of fluorine is
most sensitive to errors in effective temperature due to the relatively low
dissociation potential of HF.
Since we relied on multiple sources for atmospheric paremeters, we have no
guarantee of consistency of the temperature scale, particularly over the range
of spectral type (G5 to M0) and luminosity class (V to Iab) represented 
in our sample.  
In principle, we might expect star-to-star errors in temperature of perhaps
30~K for stars from particular literature sources, and 100~K in the full sample.
A temperature error of 100~K would lead to an error of $\sim$0.25~dex;
errors introduced from uncertainties in surface gravity, microturbulence,
and metallicity are all small, less than 0.1~dex, as shown in Table 3.  
Combining errors in quadrature suggests an uncertainty of 0.26~dex.
On the other hand, at a given metallicity, the dispersion in the measured
fluorine abundances is $\sim$0.24~dex.  The observed scatter in 
log~$\epsilon$(F) is consistent with a measurement uncertainty of 0.26~dex.

\subsection{Comparison to Previous Work}

Our fluorine abundances for thin disk stars are compared in Figure 3 (upper
panel) to other works that have included normal, thin disk dwarfs and giants
not exhibiting other evidence of contamination from AGB nucleosynthesis.  
Nearby dwarf stars are available from Recio-Blanco et al. (2012), field 
giants were included by Jorissen et al. (1992), Cunha \& Smith (2005), and
J{\"o}nsson et al. 2014b; and open cluster giants 
have been reported by Nault \& Pilachowski (2013) and Maiorca. et al (2014).
Fluorine abundances from Recio-Blanco et al., Jorissen et al., and
Cunha \& Smith all used the older excitation potential for the HF feature,
consistent with the original solar abundance of fluorine reported by Hall and 
Noyes (1969), and [F/Fe] values are computed relative to a solar abundance of
log~$\epsilon$~=~4.56.
J{\"o}nsson et al.  used similar molecular data to ours and their [F/Fe]
ratios are calculated using the new Maiorca et al. (2014) solar abundance.

Compared to other studies of the abundance of fluorine in thin disk stars in
the solar neighborhood, our much larger sample covers a wider range of 
metallicity in the thin disk (${-0.6}<[Fe/H]<{+0.3}$) and includes 
stars in the restricted temperature range 3800~$<$~T$_{eff}$~$<$~4500 K
in which the HF feature is strong enough to be easily measured.
Adopting the new Maiorca et al. (2014) solar fluorine abundance of 
log~$\epsilon$(F)~=~4.40 (Maiorca et al. used the same molecular data as are
used in this paper), the average fluorine abundance of our sample of 
nearly 40 stars with T$_{eff}$~$<$~4500 K is [F/Fe]=~+0.23~$\pm$~0.03 
(SEM).
This result compares with an average abundance [F/Fe]~=~+0.24~$\pm$~0.06 (SEM)
for the Recio-Blanco et al. (2012) sample of nine dwarf stars and an average 
of [F/Fe]~=~+0.21~$\pm$~0.03 (SEM) for the seven 
normal thin disk giants included by Jorissen et al. (1992).  
In contrast, J{\"o}nsson et al. (2014b) find [F/Fe]~=~--0.04~$\pm$~0.05 (SEM)
from their analysis of the 12~$\mu$m~HF feature in six cool K and M giants.

Each of these studies has adopted different approaches for determining the
model atmosphere parameters.  Some rely on parameters determined from optical
spectra (Recio-Blanco et al. 2012 and this paper), one relies on a
temperature vs. spectral type relation (Jorissen et al. 1992 from Smith \& 
Lambert 1990), one relies on broadband colors (Cunha \& Smith 2005) and one 
relies on optical angular diameter measurements (J{\"o}nsson et al. 2014b).
Given the sensitivity of the derived fluorine abundance to temperature
(typically 0.25 dex per 100 K temperature change; see Table 3),
temperature scale errors likely dominate the uncertainty in the average
fluorine abundance of the local thin disk.
The average fluorine abundance in the solar neighborhood remains uncertain
at the level of 0.2 dex.

A formal fit to the dependence of [F/Fe] on [Fe/H] for our sample of stars
with well-determined abundances gives 
[F/Fe]~=~--0.105[Fe/H]~+~0.20, suggesting an average fluorine abundance of 
[F/Fe]~=~+0.20 (log~$\epsilon$(F)~=~4.6) at the solar metallicity, 
somewhat higher than the Lodders et al. (2009) meteoritic abundance 
of log~$\epsilon$(F)~=~4.42 or the new Maiorca et al. (2014) solar 
photospheric abundance of log~$\epsilon$(F)~=~4.40.  
The Sun may be slightly deficient in fluorine compared to the solar 
neighborhood, but a firm conclusion awaits a self-consistent determination 
of stellar atmospheric parameters using infrared spectra.

\section{Fluorine Enrichment in the Thin Disk}

Given the variety of possible sources of fluorine in the Milky Way (SNe II,
AGB stars, WR stars, etc.), 
observations of fluorine in uncontaminated field stars may provide clues 
about the principal contributor of the element to the Galactic disk today.  
Of particular interest is the relationship between fluorine and other light
elements that are affected by proton-capture nucleosynthesis.
Studies of fluorine and oxygen in AGB stars exhibiting the products of 
the third dredge-up find a correlation between [F/Fe] and [O/Fe]
(see, for example, the fluorine abundances from Abia et al. 2010 and 
the oxygen abundances from Lambert et al. 1986 for the same stars,
as well as the
Cunha et al. 2003 samples from the Milky Way and Large Magellanic Cloud).

The average fluorine in N-type carbon stars by Abia et al. (2010) is
[F/Fe]~=~0.27~$\pm$~0.04 (SEM).  
Abia et al. note the relatively low fluorine abundances of 
the N-type stars compared to SC-type AGB stars, as well as the similarity of
their fluorine abundances to those of K and M field stars at evolutionary
phases prior to the AGB.
These abundances are suggestive that at least some N-type carbon stars 
in the Milky Way are only modestly enhanced in fluorine, if at all.
 
Observations of light elements in globular cluster stars (Carretta et al. 
2009, M\'{e}sz\'{a}ros et al. 2015) have demonstrated a substantial spread in
light element abundances attributed to proton-capture nucleosynthesis, most 
likely occuring in an earlier generation of metal-poor, intermediate-mass,
hot-bottom-burning AGB stars (although fast-rotating massive stars may also 
play a role).
Studies of fluorine in globular 
cluster giants (Cunha et al. 2003, Smith et al. 2005, Yong et al. 2008,
D'Orazi et al. 2013,  de Laverny \& Recio-Blanco 2013)
have shown both a correlation of fluorine and oxygen and an anti-correlation 
of fluorine and sodium, for clusters in the metallicity range
-1.2 $\leq$ [Fe/H] $\leq$ -0.7.  Observations of HF in cluster giants are,
however, very difficult because of the low metallicity and relatively warm
temperatures of even the coolest red giants.  The HF feature is typically
weak compared to residual features from telluric line correction
(D'Orazi et al. 2013).

Fluorine production in AGB stars occurs primarily in 2-4 solar mass stars 
according to Nomoto et al. (2013), although more massive AGB stars may also
produce excess fluorine once hot bottom burning (which destroys fluorine)
ceases (see Karakas 2010, Lugaro et al. 2012, and D'Orazi et al. 2013 for
a more complete discussion).
If present in normal stars in the Galactic thin disk, a fluorine-oxygen 
correlation and fluorine-sodium anti-correlation in thin disk stars might 
indicate some substantial contribution of AGB stars to the fluorine abundance 
in the thin disk. The question is complicated, however, by the possibility
of metallicity dependent yields from the various potential sources of fluorine
in the Milky Way.

The existence of a fluorine-oxygen correlation 
is examined explicitly for the thin disk in Fig. 4.
Oxygen abundances are available from the literature for only a few of the stars 
in the combined sample of new and previously published stars with fluorine
abundances, and are plotted as [F/O] vs. [O/H] in the upper panel of Fig. 4.  
Models of the Galactic chemical evolution of fluorine and oxygen
by Kobayashi et al. (2011) suggest that the [F/O] ratio for fluorine
production by AGB stars plus SNe (without neutrino spallation) should be 
relatively flat, with [F/O]~$\approx$~--0.15, while models that include
neutrino spallation produce a higher [F/O] ratio ([F/O]~=~+0.2), also with
no dependence of [F/O] on [O/H].
The average [F/O] abundance for the sample of available data,
including the low fluorine abundances from J{\"o}nsson et al. (2014b), is
[F/O]~=~+0.10~$\pm$~0.05~(SEM),suggesting significant contributions
from SNe II with neutrino spallation to the abundance of fluorine in the thin
disk.
The Kobayashi et al. models suggest that 2-4 solar mass AGB stars began to 
contribute substantially to the abundance of fluorine in the Galaxy at an 
oxygen abundance of [O/H]~=~--1.2, then reached a plateau at [O/H]~=~--0.5.
From the [O/Fe] versus [Fe/H] data of Bensby et al. (2004), this oxygen 
abundance corresponds to [Fe/H] of perhaps --0.9 dex. 

In Fig. 5, the fluorine abundances [F/Fe] are plotted vs. [Fe/H], this time
overlain with the Kobayashi et al. (2011) predictions for three nucleosynthesis
models: standard supernovae (including SNe II, hypernovae, and SNIa), supernovae
plus contributions from AGB stars, and, finally, with additional contributions
from the $\nu$-process in core-collapse supernovae. 
The reader is cautioned, however, that the Kobayashi et al. models do not 
include possible contributions from WR stars.
Since oxygen abundances are available for so few stars in our sample,
[F/Fe] and [Fe/H] ratios have been obtained from the Kobayashi et al. models
by assuming a linear correlation between the [O/Fe] and [Fe/H] ratios
for thin disk stars in this metallicity range from Bensby et al.
(2004), combined with the [O/H] and [F/O] model predictions.

We note first that, although the scatter in our measurements is large, 
the observed abundance of fluorine in the local thin disk matches well the 
predictions of Kobayashi et al. (2011) for models including neutrino spallation, and
is well above the prediction for production dominated by AGB stars.
The average abundance of fluorine at solar metallicity for our data
is [F/Fe]~=~+0.2~$\pm$~0.03 (SEM), while
the models predict [F/Fe]~=~0.18 for the case of a neutrino explosion energy
of 3~x~10$^{53}$~erg, and higher for higher neutrino explosion energies.

The slope of the dependence of [F/Fe] on [Fe/H] is also consistent with the
models including neutrino spallation within our estimated uncertainties.  
For our data in the range --0.6~$<$~[Fe/H]~$<$~+0.3, we obtain a slope of 
--0.105 ~$\pm$~0.14 per dex, 
compared to the model prediction of --0.18 per dex.
In contrast, Kobayashi et al.'s (2011) model for fluorine production 
from AGB stars and SNe II without neutrino spallation indicates
that the [F/Fe] ratio should increase by 0.2 dex over the same metallicity
range. Such an increase is inconsistent with our observations of a flat or 
slowly decreasing [F/Fe] ratio with metallicity at the solar metallicity. 
Measurements of the slope of the [F/Fe] vs [Fe/H] dependence are independent
of any systematic errors in temperature scale. 
The uncertainty on the measured slope can be reduced with better
determinations of the atmospheric parameters, particularly the 
stellar effective temperatures. 
 
Both the observed decrease of [F/Fe] with [Fe/H] and the average 
value of [F/Fe] at the solar metallicity appear to support significant 
contributions of fluorine from supernovae through neutrino spallation
in SNe II.  Our conclusion is in contrast to the conclusions of both
Recio-Blanco et al. (2012) and J{\"o}nsson et al. (2014b).
Recio-Blanco et al. found, based in part on a possible correlation of [F/Fe]
with s-process abundances and the apparent absence of a correlation of [F/Fe]
with alpha elements, that AGB stars were the likely producers of fluorine in the
Galactic disk.  However, their analysis included only nine relatively warm 
main sequence stars (Teff~$>$~4500 K) with quite weak HF features, and 
their statistical correlations between fluorine and the s-process and alpha
elements are low ($\approx$0.5), as they note.
Given their small sample size and the amplitude of any trends compared with
their uncertainties, the Recio-Blanco et al. sample does not provide
compelling evidence for AGB-dominated production of fluorine.

J{\"o}nsson et al. (2014b), conclude that only AGB production of fluorine 
is needed to account for the abundance of fluorine in
their sample of six cool, thin disk giants. 
Their conclusion is based on the relatively low derived abundance of fluorine 
in five of their six thin-disk giants, comparing to model predications of both 
[F/Fe] and [F/O]. 
Given the temperature sensitivity of HF due to its low dissociation energy, 
combined with the extrapolation of interferometric diameters from optical 
wavelengths to determine temperatures in the infrared, systematic errors may 
affect their fluorine abundances at the level of 0.2 dex, which might alter
their conclusions about its origin.

In the end, the source of fluorine in the Galactic thin disk remains
relatively unconstrained by the observations, particularly given the diversity
of potential sources of the element.  Systematic errors among the 
temperature scales of the abundance determinations remain a significant
source of uncertainty, as do the model predictions in light of the recent
revision of the solar abundance of fluorine downward (Maiorca et al. 2014).
The lack of oxygen abundance measurements in the same stars also limits our
ability to constrain fluorine production models.
Production in AGB stars remains a strong contender because of the observation
of enhancements in fluorine in some AGB stars, but those enhancements, while
interesting, are not themselves compelling evidence for a dominant role for 
AGB stars in Galactic chemical evolution.  The question is further clouded
by the discovery of strong fluorine enhancements in some CEMP stars 
(Schuler et al. 2007 and Lucatello et al 2011), and of strong fluorine 
enhancements in AGB stars in the metal-poor Large Magellanic Cloud 
(Abia et al. 2011), forcing metallicity-dependent
yields into the mix of production scenarios.

\section{Summary}

Fluorine 
abundances have been determined for several dozen
dwarf and giant stars identified as belonging to the Galactic thin disk to 
characterize the abundance of fluorine and its possible nucleosynthesis
sources.
We enumerate our basic conclusions below.
\begin{itemize}
\item{Studies of normal stars in the Milky Way thin disk provide an
opportunity to constrain models of fluorine nucleosynthesis.  We provide
the largest sample of normal stars yet analyzed for fluorine abundances.}
\item{In the metallicity range --0.6~$\leq$~[Fe/H]~$\leq$~+0.3, we find an
average [F/Fe] ratio of +0.22~$\pm$~0.03 (SEM), although systematic errors in
the temperature scale may contribute an uncertainty of 0.2 dex in comparing
different studies.}
\item{In the same metallicity range, the [F/Fe] ratio is constant or slowly
declining with metallicity, within our estimated uncertainties.}
\item{The abundance of fluorine compared to oxygen in the thin disk is slightly
high, typically [F/O]~=~+0.10~$\pm$~0.05.}
\item{The observed [F/Fe] abundance in the thin disk, the measured
slope of the [F/Fe] versus [Fe/H] relation, and the relatively high [F/O]
ratio all provide evidence for contributions from the neutrino process in 
core-collapse supernovae to the abundance of fluorine in the thin disk and are
inconsistent with fluorine production only in AGB stars.}
\item{The presence or absence of correlations or anti-correlations of [F/Fe] 
with [O/Fe] or [Na/Fe] may tell us about the contributions of AGB stars
and/or fast-rotating massive stars toward the production of fluorine through 
proton-capture processing the Milky Way disk, but more accurate measurements
of the [F/Fe] ratio are needed, as are measurements of the oxygen abundance
in the same stars.}
\item{The origin of fluorine in the Galactic disk remains relatively
unconstrained by observations, both because of possible systematic errors
in temperature scale and because of insufficient data for other light 
elements in the same stars.}
\end{itemize}

\acknowledgments

We are grateful to the Kitt Peak National Observatory and 
particularly to Dick Joyce and Krissy Reetz 
for their assistance at the start of the observing run.
This research has made use of the NASA Astrophysics Data System 
Bibliographic Services, the HITRAN database operated by the Center 
for Astrophysics, and the WEBDA database and Vienna Atomic
Line Database, both operated at the Institute
for Astronomy of the University of Vienna.
This research has made use of the SIMBAD database, operated at CDS,
Strasbourg, France.
We thank Eric Ost for implementing the model atmosphere 
interpolation code.
Finally, we thank an anonymous referee for detailed comments on the 
manuscript.
C.A.P. acknowledges the generosity of the Kirkwood Research Fund at 
Indiana University.

\clearpage

\setlength{\parindent}{0cm}

References

Abia, C. Recio-Blanco, A., de Laverny, P.,
Chritallo, S., Dom\'{i}nguez, I., \& Straniero, O. 2009, \apj, 694, 971

Abia, C., Cunha, K., Cristallo, S.,
de Laverny, P., Dom\'{i}nguez, I., Eriksson, K., Gialanella, L.,
Hinkle, K., Imbriani, G., Recio-Blanco, A., Smith, V. V., Straniero, O.,
\& Wahlin, R. 2010, \apj, 715, L94

Abia, C., Cunha, K., Cristallo, S.,
De Laverny, P., Dom\'{i}nguez, I., Recio-Blanco, A., Smith, V. V., 
\& Straniero, O. 2011, \apjl, 737, L8

Allende Prieto, C., Barklem, P. S.,
Lambert, D. L., \& Cunha, K. 2004, \aap, 420, 183

Asplund, M. 2005, ARAA, 43, 481

Balachandran, S. C., 
Fekel, F. C., Henry, G. W., \& Uitenbroek, H. 2000, \apj, 542, 978

Bensby, T.; Feltzing, S.; \& Lundstr\"{o}m, I. 2004, \aap, 415, 155

Carretta, E., Bragaglia, A., Gratton, R. G.,
Lucatello, S., Catanzaro, G., Leone, F., Bellazzini, M., Claudi, R., 
D'Orazi, V., Momany, Y., Ortonaly, S., Pancino, E., Piotto, G.,
Recio-Blanco, A., Sabbi, E. 2009, \aap, 505, 117

Casagrande, L., Sch\"{o}nrich, R.,
Asplund, M., Cassisi, S., Ram\'{i}rez, I., Mel\'{e}ndez, J.,
Bensby, T., \& Feltzing, S. 2011, \aap, 530, 138

Cristallo, S., Straniero, O.,
Gallino, R., Piersanti, L., Dom\'{i}nguez, I., \& Lederer, M. T. 2009,
\apj, 696, 797

Clayton, D. 2003, Handbook of Isotopes in the
Cosmos Hydrogen to Gallium (Cambridge: Cambridge University Press)

Cunha, K.; Smith, V. V.; Lambert, D. 
L.; Hinkle, K. H. 2003, \aj, 126, 1305

Cunha, K., \& Smith, V. V. 2005, \apj, 626, 425

de Laverny, P. \& Recio-Blanco, A. 2013, \aap, 555, 121

D'Orazi, V.; Lucatello, S.; 
Lugaro, M.; Gratton, R. G.; Angelou, G.; Bragaglia, A.; Carretta, E.; 
Alves-Brito, A.; Ivans, I. I.; Masseron, T.; \&Mucciarelli, A. 2013, 
\apj, 763, 22

Gallino, R., Bisterzo, S., Cristallo, S.,
\& Straniero, O. 2010, Mem. Soc. stron. Ital., 81, 998

Garc\'{i}a-Hern\'{a}ndez,
D. A., Garc\'{i}a-Lario, P, Plez, B., et al. 2006, Science, 314, 1751

Garc\'{i}a-Hern\'{a}ndez,
D. A., Garc\'{i}a-Lario, P, Plez, B., et al. 2007, \aap, 462, 711

Goorvitch, D. 1994, \apjs, 95, 535.

Gratton, R. G., Carretta, E., \&
Bragaglia, A. 2012, Astron. Astrophys. Rev. 20. 50

Gustafsson B., Edvardsson B., Eriksson K.,
 Jorgensen U.G., \& Nordlund, \AA., \& Plez B. 2008, \aap  486, 951.

Hall, D. N. B. \& Noyes, R. W. 1969, \apjl, 4, 13

Hekker, S., \& Mel\'{e}ndez 2007, \aap, 475, 1003

Hinkle, K.H., Cuberly, R., Gaughan, N., 
Heynssens, J., Joyce, R.R. Ridgway, S.T., Schmitt, P., and Simmons, J.E. 1998, Proc. SPIE 3354, 810

Jeffery, S. C., Karakas, A. I., \&
Saio, H. 2011, \mnras, 414, 3599

Johnson, D. R. H., \& Soderblom, D. R. 1987, \aj, 93, 864

J{\"o}nsson, H., Ryde, N.,
Harper, G. M., Cunha, K., Schultheis, M., Eriksson, K., Kobayashi, C.,
Smith, V. V., \& Zoccali, M. 2014a, \aap, 564, A122

J{\"o}nsson, H., Ryde, N.,
Harper, G. M., Richter, M. J., \& Hinkle, K. H. 2014b, \apjl, 789, L41

Jorissen, A.; Smith, V. V.; \& 
Lambert, D. L. 1992, \aap, 261, 164

Karakas, A. I., Lee, H. Y., Lugaro, M., 
G{\"o}rres, J. \& Wiescher, M. 2008, \apj, 676, 1254

Karakas, A. I. 2010, \mnras, 403, 1413

Kobayashi, C., Izutani, N., Karakas, A. I.,
Yoshida, T., Yong, D., \& Umeda, H. 2011, \apjl, 739, L57

Kotoneva, E., Shi, J. R., \& Zhao, G. 2006, \aap, 454, 833

Kupka F., Ryabchikova T.A., Piskunov N.E., 
Stempels H.C., \& Weiss W.W. 2000, BaltA, 9, 590

Lambert, D. L., Gustafsson, B.,
Eriksson, K., \& Hinkle K. H. 1986, \apj, 62, 373

Li, H. N., Ludwig, H.-G., Caffau, E.,
Christlieb, N. \& Zhao, G. 2013, \apj, 765, 51

Lodders, K., Palme, H., \& Gail, H.-P.
2009, Landolt-B{\"o}rnstein, 44

Lucatello, S.; Masseron, T.; 
Johnson, J. A.; Pignatari, M.; Herwig, F. 2011, \apjl, 729, 40

Luck, R. E., \& Bond, H.E. 1980,\aj, 241, 218

Luck, R. E. \& Heiter, U. 2006, \aj, 131, 3069

Lugaro, M., Ugalde, C., Karakas, A. I.,
G\"{o}rres, J., Weischer, M., Lattanzio, J., C., \& Cannon, R. C. 2004,
\apj, 615, 934

Lugaro, M., Karakas, A. I., Stancliffe,
R. J., \& Rijs, C. 2012, \apj, 747:2

Lyubimkov, L. S., Lambert, D. L.,
Rostopchin, S. I., Rachkovskaya, T., \& Poklad, D. B. 2010, \mnras, 402, 1369

Maiorca, E., Uitenbroek, H., 
Uttenthaler, S., Randich, S., Musso, M., \& Magrini, L. 2014, \apj, 788, 149

Masseron, T. 2006, PhD thesis, Obs. de Paris

M\'{e}sz\'{a}ros, Sz., Martell, S. L.,
Shetrone, M. et L. 2015, \aj, 149, 153

McWilliam, A. 1990, \apjs, 74, 1075

Mel\'{e}ndez, J., Asplund, M.,
Alves-Brito, A., Cunha, K., Barbuy, B., Bessell, M. S., Chappini, C.,
Freeman, K. C., Ram\'{i}rez, I., Smith, V. V., \& Yong, D. 2008,
\aap, 484, L21

Meynet, G. \& Arnould, M. 
1993,  in Nuclei in the Cosmos II, ed. K\"{a}ppeler, F., Wisshak, E.,
(Bristol, IOP), p. 503

Meynet, G. \& Arnould, M. 2000, \aap, 355, 176

Mishenina, T. V., Bienaym\'{e}, O.,
Gorbaneva, T. I., Charbonnel, C., Soubiran, C., Korotin, S. A., \&
Fovtyukh, V. V. 2006, \aap, 456, 1109

Mishenina, T. V., Soubiran, C.,
Bienaym\'{e}, O., Korotin, S. A., Belik, S. I., Usenko, I. A., \&
Kovtyukh, V. V. 2008, \aap, 489, 923

Nault, K. A. \& Pilachowski, C. A. 2013, \aj,  146, 153

Nomoto, K.; Kobayashi, C., \&
Tominaga, N. 2013, \araa, 51, 457

Palacios, A., Arnould, M., \& Meynet, G. 2005, \aap, 443, 243

Pandey, G. 2006, \apj, 648, L143

Pandey, G., Lambert, D. L., \& Rao, N. K. 2008, \apj, 674, 1068

Prugniel, Ph., Vauglin, I', \& Koleva, M. 2011, \aap, 531, A165

Ram\'{i}rez, I.\& Allende Prieto, C. 2011, \apj, 743:135

Ram\'{i}rez, I., Allende Prieto, C., \& Lambert, D. L. 2013, \apj, 764:78

Ram\'{i}rez, S. V.,
Sellgren, K., Carr, J. S., Balachandran, S. C., Blum, R., Terndrup, D. M.,
\& Steed, A. 2000, \apj, 537, 205

Recio-Blanco, A., de Laverny, P.,
Worley, C., Santos, N. C., Melo, C. \& Israelian, G. 2012, \aap, 538, 117

Rothman, L. S., Gorden I. E., Babikov, Y., et al. 2013, JQSRT, 130, 4

Schuler, S. C., Cuhna, K., Smith, V. V.,
Sirvarnai, T., Beers, T. C., \& Lee, Y. S. 2007, \apj, 667, 81 

Smith, V. V. \& Lambert, D. 1990, \apjs, 72, 387

Smith, V. V., Cunha, K., Ivans, I. I.,
Lattanzio, J. C., Campbell, S., \& Hinkle, K. H. 2005, \apj, 633, 392

Sneden, C. 1973, \apj, 184, 839

Sneden, C., Cowan, J. J, Lawler, J. E., et al. 2003, \apj, 591, 936

Soubiran, C., Le Campion, J.-F.,
Cayrel de Strobel, G., \& Caillo, A. 2010, \aap, 515, A111

Sousa, S. G., Santos, N. C., Mayor, M.,
Udry, S., Casagrande, L., Israelian, G., Pepe, F. Wueloz, D. \&
Monteiro, M. J. P. F. G. 2008, \aap, 487, 373

Spite, M., Cayrel, R., Plez, B., 
Hill, V., Spite, F., Depagne, E., Francios, P.. Bonifacio, P., Barbuy, B.,
Beers, T., Andersen, J., Molaro, P., Nordstr\"{o}m, B., \& Primas, F. 2005,
\aap, 430, 655

Takeda, Y., Sato, B., \& Murata, D.
 2008, \pasj, 60, 781

Tinsley, B. 1980, FCPh, 5, 287

Valenti, J. A., \& Fischer, D. A. 2005, \apjs, 159, 141

Ventura, P. \& D'Antona, F.,
Mazzitelli, I, \& Gratton, R. 2008, \apj, 550, L65

Ventura, P. \& D'Antona, F. 2001, \aap, 479, 805

Wallace, L. \& Livingston, W. 1992, N.S.O. Technical Report \#92-001.

Woosley, S. E.; Hartmann, D. H.; 
Hoffman, R. D.; \& Haxton, W. C. 1990, \apj, 356, 272

Woosley, S. E., Heger, A., \&
Weaver, T. A. 2002, Review of Modern Physics, 74, 1015

Yong, D., Mel\'{e}ndez, J., Cunha, K.,
Karakas, A. I., Norris, J. E., \& Smith, V. V. 2008, \apj, 689, 1020

\clearpage


\begin{figure}
\epsscale{.80}
\plotone{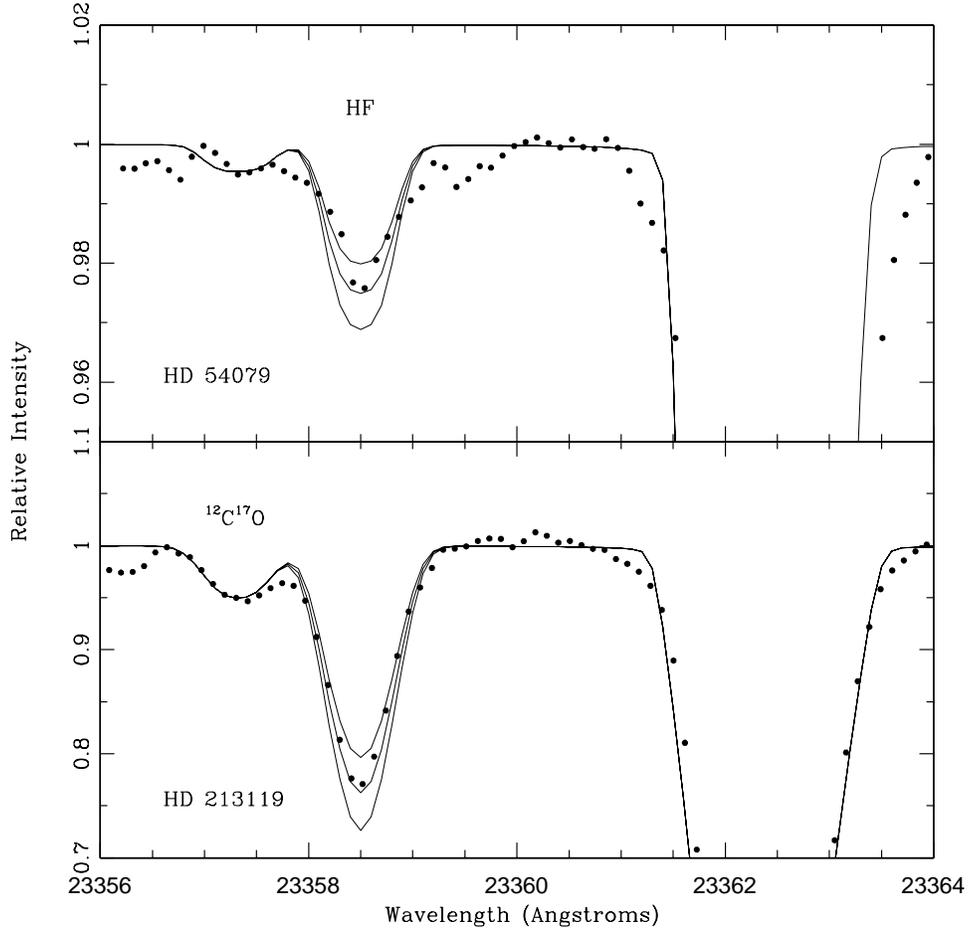}
\caption{HF spectra of two thin disk stars are shown together with   
calculated synthetic spectra.  The upper panel is the spectrum of HD 54079,
a giant with a temperature of 4450 K and [Fe/H] = --0.45.  It is shown with 
synthetic spectra for log A(F) = 3.98, 4.08. and 4.08.  The lower panel
is the spectrum of HD23119, a giant with a temperature of 4090 K and 
[Fe/H] = --0.50.  It is shown with synthetic spectra for log A(F) = 4.24,
4.34, and 4.44.  The $^{12}$C/$^{17}$O feature adjacent to HF has been noted in
the lower panel.}
\label{fig1_spec}
\end{figure}

\begin{figure}
\epsscale{.80}
\plotone{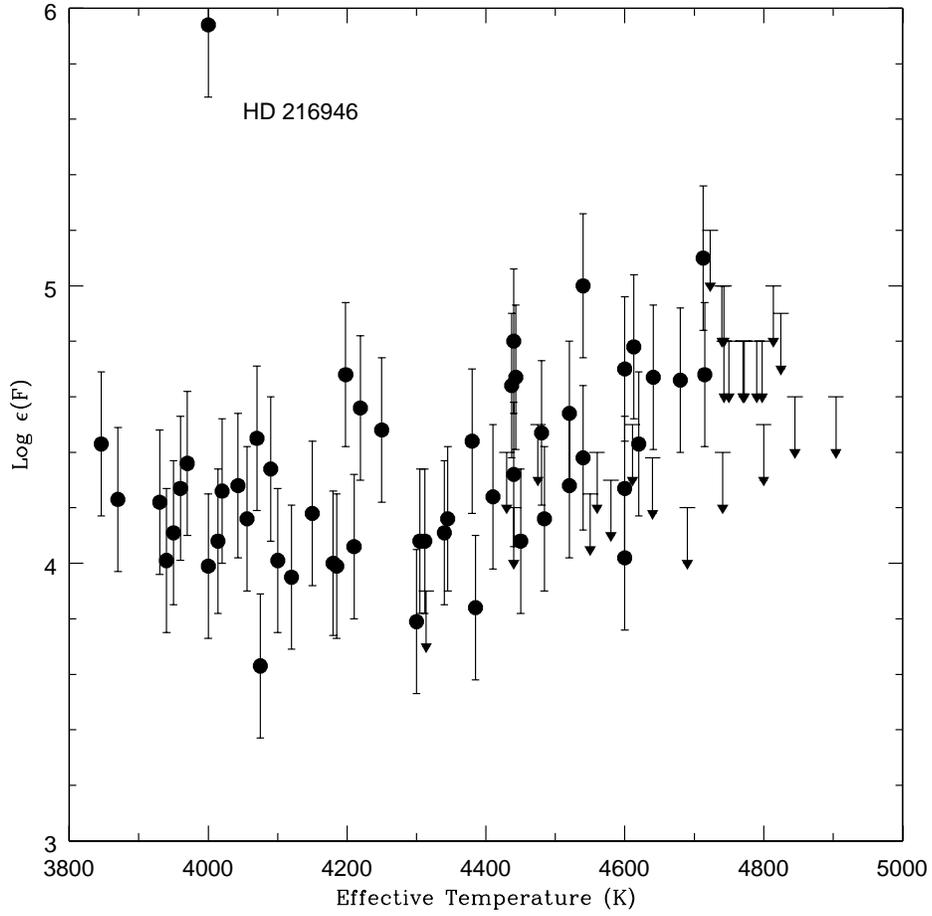}
\caption{Derived abundance of log $\epsilon$(F) versus 
effective temperature. 
The HF feature becomes undetectable at temperatures above 4700 K.  For
temperatures 4500 $<$ T$_{eff}$ $<$ 4700, HF is detectable only if the abundance
is high. 
The M0Iab supergiant HD 216946 (HR 8726) is labelled.}
\label{fig2_teff}
\end{figure}

\begin{figure}
\epsscale{.80}
\plotone{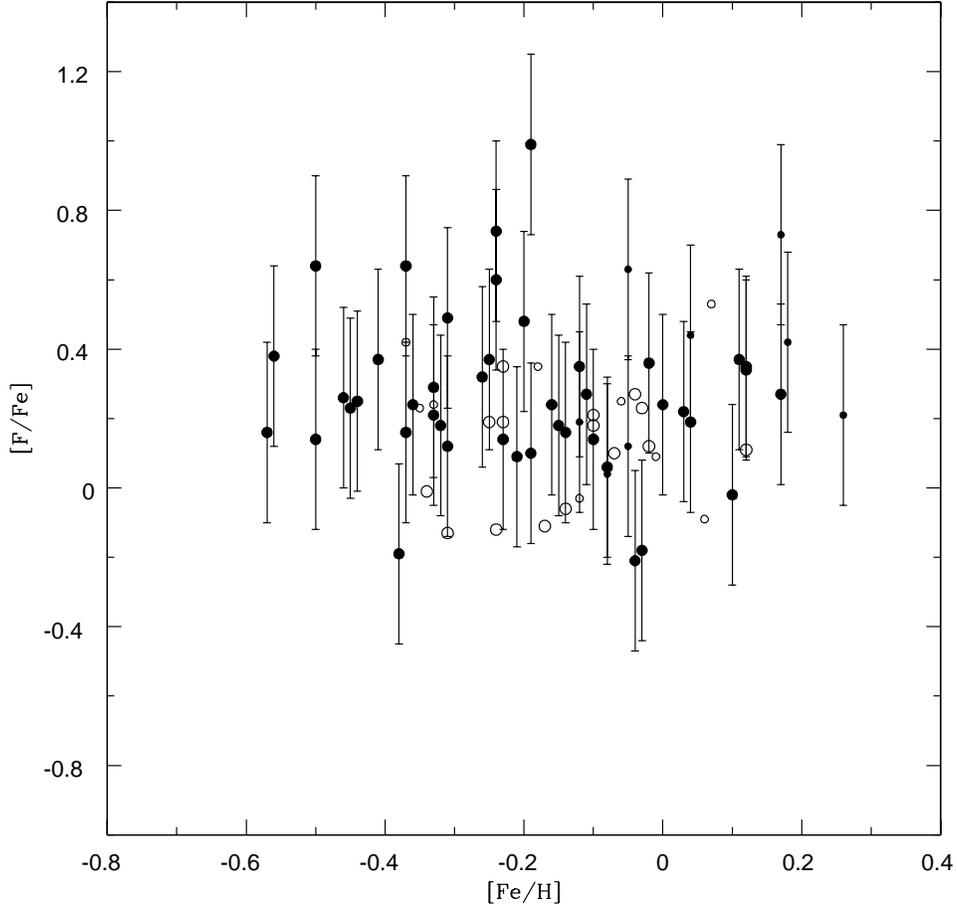}
\caption{Derived abundance of fluorine [F/Fe] versus [Fe/H] 
plotted together with values from the literature.  
Stars with temperatures above 4500 K
have been omitted, since many are upper limits due to the high temperature.
Filled symbols represent stars
reported here.  Large filled circles are giants and small filled circles
are dwarfs from our measurements.  
Literature measurements from Jorissen et al. (1992), Recio-Blanco et al. (2012),
Nault \& Pilachowski (2013), and J{\"o}nsson et al. (2014b)
are shown as open circles.}  
\label{fig3_f}
\end{figure}

\begin{figure}
\epsscale{.80}
\plotone{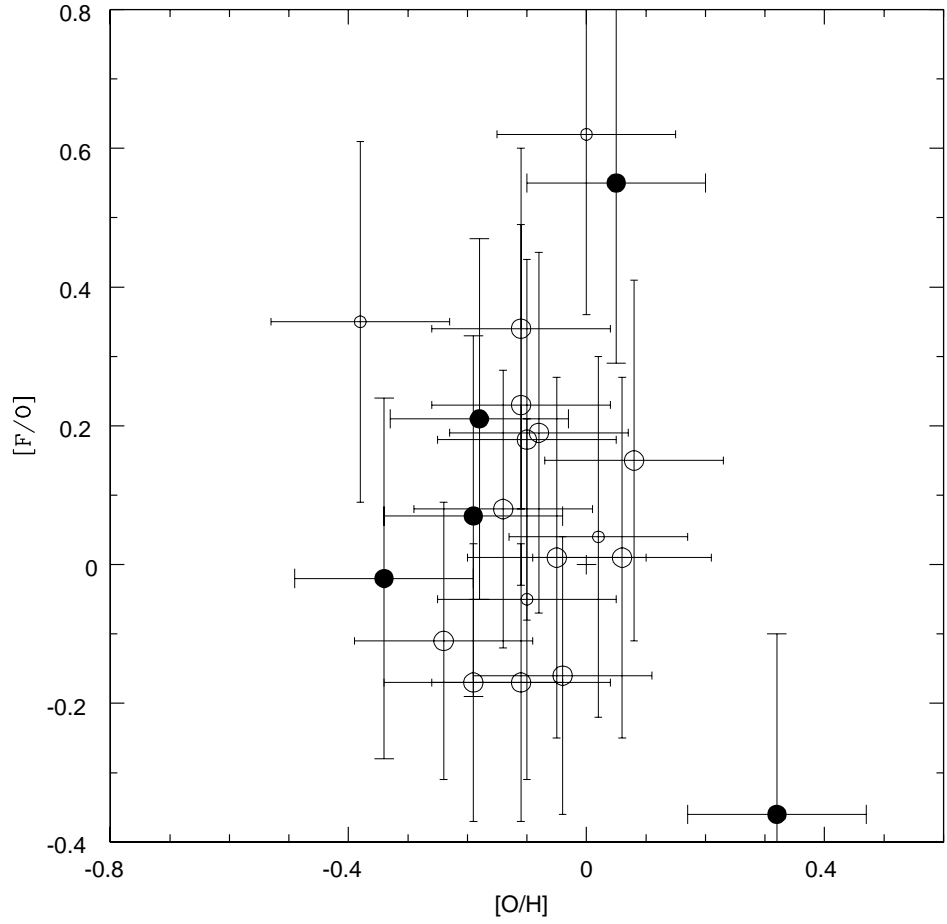}
\caption{[F/O] versus [O/H] for thin disk stars, 
including measurements from the literature.  
Symbols are as defined in Fig. 3.}
\label{fig4_cor}
\end{figure}

\begin{figure}
\epsscale{.80}
\plotone{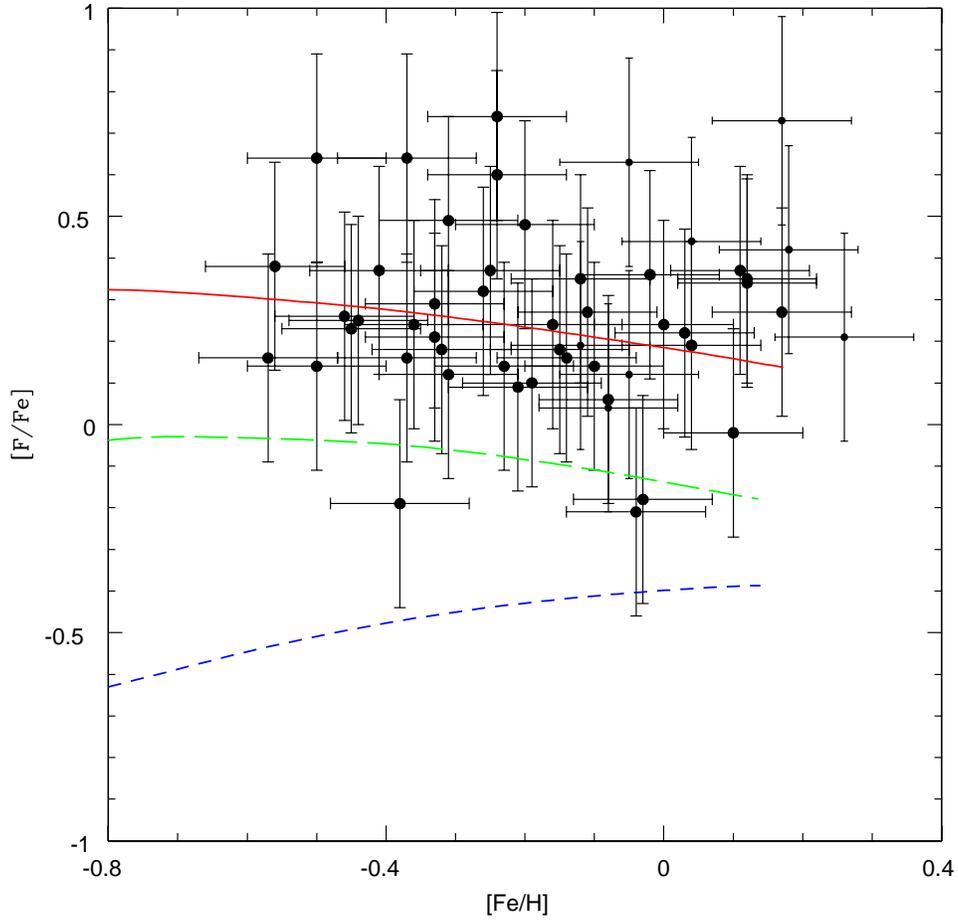}
\caption{[F/Fe] versus [Fe/H] for thin disk stars.  Chemical evolution models 
from Kobayashi et al. (2011) are shown for AGB production (small-dashed blue
curve), SN plus AGB production (large-dashed green curve), and SN production 
with neutrino spallation plus contributions from AGB stars (solid red curve).}
\label{fig5_th}
\end{figure}






\clearpage

\begin{deluxetable}{rrllcrrr}
\tabletypesize{\scriptsize}
\tablecaption{Stars Observed\label{table_log}}
\tablewidth{0pt}
\tablehead{
& & & \colhead{Spec.} &  \colhead{UT} & & \colhead{Exp.} & \colhead{S/N} \\
\colhead {HD} & \colhead{HIP} &  \colhead{ Alt.} & \colhead{Type} &
\colhead{Date} & \colhead{K Mag.} & \colhead{Time (s)} & \colhead{Ratio} 
}
\startdata
6 & 417 & HR 2 & G9III: & 3 Dec. 2012 & 3.908 & 4 x 300 & 175 \\
5268 & 4257 & 21 Cet & G5IV & 3 Dec. 2012 & 3.91 & 4 x 300 & 300\\
5437 & 4371 & 22 Cet & K4III & 3 Dec. 2012 & 1.826 & 4 x 30 & 200\\
6805 & 5364 & eta Cet & K1III & 5 Dec. 2012 & 0.875 & 4 x 30 & 150\\
7578 & 5936 & HR 371 & K1III & 2 Dec. 2012 & 3.371 & 4 x 400 & 130\\
8207 & 6411 & ksi And & K0III & 2 Dec. 2012 & 2.554 & 4 x 150 & 200\\
9408 & 7294 & chi Cas & G9IIIb & 2 Dec. 2012 & 2.311 & 4 x 120 & 180\\
9900 & 7617 & HR 461 & G5Iab: & 2 Dec. 2012 & 2.391 & 4 x 120 & 180\\
9927 & 7607 & ups Per & K3III & 2 Dec. 2012 & 0.649 & 3 x 30 & 200\\
10853 & 8275 & \nodata & K3.5V & 4 Dec. 2012 & 6.319 & 4 x 900 & 200\\
11949 & 9222 & 3 Per & K0IV & 1 Dec. 2012 & 3.03 & 4 x 90 & 210\\
12929 & 9884 & alf Ari & K2III & 1 Dec. 2012 & -0.783 & 4 x 5 & 240\\
13520 & 10340 & b And & K4III & 29 Nov.~2012 & 1.329 & 4 x 100 & 250\\
13789 & 10416 & LTT 1147 & K3.5V & 1 Dec. 2012 & 6.015 & 4 x 900 & 120\\
15755 & 11840 & HR 738 & K0III & 5 Dec. 2012 & 3.233 & 4 x 300 & 192\\
17660 & 13258 & LHS 1453 & K4.5V & 4 Dec. 2012 & 6.128 & 4 x 900 & 170\\
18449 & 13905 & 24 Per & K2III & 5 Dec. 2012 & 2.103 & 4 x 90 & 190\\
19270 & 14439 & HR 931 & K3III & 3 Dec. 2012 & 3.263 & 4 x 300 & 195\\
20644 & 15549 & HR 999 & K4III & 1 Dec. 2012 & 0.877 & 4 x 20 & 200\\
21017 & 15861 & 64 Ari & K4III & 5 Dec. 2012 & 2.888 & 4 x 300 & 200\\
26162 & 19385 & \nodata & F8V & 3 Dec. 2012 & 6.944 & 4 x 300 & 230\\
26546 & 19641 & HR 1295 & K0III & 3 Dec. 2012 & 3.641 & 4 x 300 & 250\\
27382 & 20250 & phi Tau & K1III & 2 Dec. 2012 & 2.292 & 4 x 100 & 280\\
29139 & 21421 & alf Tau & K5III & 2 Dec. 2012 & -3.04 & 4 x 1 & 200\\
29697 & 21818 & V834 Tau & K3V & 4 Dec. 2012 & 5.146 & 4 x 600 & 150\\
30454 & 22393 & HR 1529 & K1III & 3 Dec. 2012 & 2.482 & 4 x 100 & 240\\
30504 & 22453 & 1 Aur & K4III & 30 Nov.~2012 & 1.307 & 4 x 30 & 200\\
30834 & 22678 & 2 Aur & K3III & 30 Nov.~2012 & 1.389 & 4 x 30 & 240\\
31539 & 23043 & HR 1585 & K1III & 30 Nov.~2012 & 2.669 & 4 x 100 & 130\\
32147 & 23311 & LHS 200 & K3V & 2 Dec. 2012 & 3.706 & 4 x 300 & 140\\
33554 & 24197 & HR 1684 & K5III & 3 Dec. 2012 & 1.732 & 4 x 30 & 280\\
34043 & 24450 & HR 1709 & K4III & 3 Dec. 2012 & 2.537 & 4 x 100 & 240\\
34334 & 24727 & 16 Aur & K2.5IIIb & 30 Nov.~2012 & 1.386 & 4 x 30 & 170\\
35620 & 25541 & phi Aur & K3IIICN & 3 Dec. 2012 & 2.104 & 4 x 90 & 200\\
36003 & 25623 & LHS 1763 & K5V & 4 Dec. 2012 & 4.88 & 4 x 600 & 150\\
37984 & 26885 & b Ori & K1III & 2 Dec. 2012 & 2.212 & 4 x 100 & 290\\
39019 & 27581 & 135 Tau & G9III: & 3 Dec. 2012 & 3.383 & 4 x 300 & 300\\
39118 & 27588 & HR 2024 & G8III & 29 Nov.~2012 & 3.337 & 4 x300 & 350\\
39715 & 27918 & LHS 1798 & K3V & 3 Dec. 2012 & 6.352 & 4 x 900 & 170\\
40657 & 28413 & HR 2113 & K1.5III & 30 Nov.~2012 & 1.649 & 4 x 60 & 150\\
43039 & 29696 & kap Aur & G8.5IIIb & 5 Dec. 2012 & 1.712 & 6 x 200 & 300\\
47174 & 31832 & 50 Aur & K3Iab: & 5 Dec. 2012 & 1.931 & 4 x 200 & 100\\
47752 & 32010 & G 109-20 & K3.5V & 4 Dec. 2012 & 5.546 & 4 x 900 & 300\\
49293 & 32578 & 18 Mon & K0IIIa & 5 Dec. 2012 & 1.849 & 5 x 200 & 120\\
52556 & 33914 & HR 2632 & K1III: & 1 Dec. 2012 & 3.105 & 4 x 300 & 100\\
52960 & 34033 & HR 2649 & K3III & 1 Dec. 2012 & 2.044 & 4 x 120 & 100\\
54079 & 34387 & HR 2682 & K0III: & 29 Nov.~2012 & 3.036 & 4 x 200 & 200\\
54716 & 34752 & 63 Aur & K3.5III & 1 Dec. 2012 & 1.605 & 4 x 45 & 290\\
54719 & 34693 & tau Gem & K2III & 1 Dec. 2012 & 1.681 & 4 x 45 & 150\\
58207 & 36046 & iot Gem & G9IIIb & 5 Dec. 2012 & 1.562 & 4 x 100 & 170\\
58972 & 36284 & gam CMi & K3III & 1 Dec.2012 & 0.993 & 4 x 30 & 230\\
60522 & 36962 & ups Gem & M0III & 30 Nov.~2012 & 0.232 & 4 x 10 & 175\\
63752 & 38253 & HR 3047 & K3III & 30 Nov.~2012 & 2.414 & 4 x 100 & 200\\
65277 & 38931 & G 113-7 & K3V & 4 Dec. 2012 & 5.511 & 4 x 900 & 200\\
65953 & 39211 & V645 Mon & K4III & 30 Nov.~2012 & 1.164 & 4 x 30 & 200\\
66141 & 39311 & HR 3145 & K2III & 30 Nov.~2012 & 1.447 & 4 x 30 & 250\\
87883 & 49699 & \nodata & K0V & 3 Dec. 2012 & 5.314 & 4 x 600 & 200\\
88320 & \nodata & \nodata & F2III & 3 Dec. 2012 & 9.038 & 4 x 300 & 200\\
209747 & 109068 & 22 Peg & K4III & 3 Dec. 2012 & 1.672 & 4 x 30 & 150\\
209945 & 109102 & HR 8424 & K5III & 3 Dec. 2012 & 1.295 & 20 x 30 & 200\\
210354 & 109354 & 27 Peg & G6III: & 3 Dec. 2012 & 3.169 & 4 x 90 & 130\\
210762 & 109602 & HR 8466 & K0 & 3 Dec. 2012 & 2.567 & 4 x90 & 190\\
211075 & 109793 & \nodata & K2 & 1 Dec. 2012 & 5.217 & 4 x 400 & 170\\
213119 & 110986 & 36 Peg & K5III & 1 Dec. 2012 & 1.882 & 4 x 30 & 210\\
214868 & 111944 & 11 Lac & K2III & 4 Dec. 2012 & 1.668 & 4 x 60 & 200\\
214995 & 112067 & \nodata & K2 III & 4 Dec. 2012 & 6.339 & 4 x 300 & 180\\
215182 & 112158 & eta Peg & G2II-III & 5 Dec. 2012 & 1.018 & 7 x 15 & 200\\
215665 & 112440 & lam Peg & G8Iab: & 4 Dec. 2012 & 1.508 & 4 x 60 & 250\\
216174 & 112731 & HR 8688 & K1III & 1 Dec. 2012 & 2.625 & 4 x 60 & 200\\
216646 & 113084 & HR 8712 & K0III & 4 Dec. 2012 & 3.514 & 4 x 300 & 250\\
216946 & 113288 & V424 Lac & M0Iab: & 5 Dec. 2012 & 0.724 & 4 x 60 & 180\\
218031 & 113919 & LTT 16772 & K0IIIb & 5 Dec. 2012 & 2.154 & 4 x 300 & 100\\
219134 & 114622 & HR 8832 & K3V & 4 Dec. 2012 & 3.26 & 4 x 300 & 300\\
220009 & 115227 & 7 Psc & K2III & 2 Dec. 2012 & 1.993 & 4 x 90 & 280\\
222107 & 116584 & lam And & G8III & 5 Dec. 2012 & 1.466 & 4 x 80 & 300\\
223559 & 117567 & HR 9029 & K4III & 3 Dec. 2012 & 2.058 & 4 x 90 & 100\\
223807 & 117756 & HR 9040 & K0III & 5 Dec. 2012 & 3.186 & 4 x 300 & 100\\
225212 & 355 & 3 Cet & K3Iab: & 3 Dec. 2012 & 1.398 & 4 x 60 & 250\\
233517 & \nodata & \nodata & K2 & 30 Nov. 2012 & 6.637 & 4 x 900 & 200
\enddata


\end{deluxetable}

\clearpage

\begin{deluxetable}{lrrcrrrr}
\tabletypesize{\scriptsize}
\tablecaption{Stellar Parameters and Fluorine Abundances\label{table_pars}}
\tablewidth{0pt}
\tablehead{
 & \colhead{$T_{eff}$} & & \colhead{$\xi$} &  &  & \colhead{F EQW} &  \\
\colhead{Star} & \colhead{(K)} & \colhead{Log $g$} & \colhead{(km s$^{-1}$)}
& \colhead{[Fe/H]} & \colhead{Ref} & \colhead{(m\AA\ )} & 
\colhead{Log $\epsilon$($F$)} 
}
\startdata
HD 6 & 4580 & 2.70 & 2.1 & 0.01 & a & $<$10 & $<$4.3  \\
HD 5268 & 4904 & 2.35 & 2.0 & --0.57 & b & $<$9 & $<$4.6  \\
HD 5437 & 3940 & 1.67 & 2.1 & --0.31 & a & 140 & 4.01  \\
HD 6805 & 4520 & 2.80 & 2.1 & --0.03 & a & 15.8 & 4.28  \\
HD 7578 & 4680 & 2.50 & 1.4 & 0.12 & c & 10.5 & 4.66  \\
HD 8207 & 4750 & 2.75 & 1.5 & 0.27 & c & $<$20 & $<$4.8 \\
HD 9408 & 4814 & 2.46 & 2.0 & --0.31 & b & $<$19 & $<$4.5  \\
HD 9900 & 4430 & 1.18 & 3.5 & --0.05 & d & $<$25 & $<$4.4  \\
HD 9927 & 4380 & 2.34 & 2.3 & 0.00 & a & 37.7 & 4.44  \\
HD 10853 & 4600 & 4.65 & 0.8 & --0.12 & c & 10.5 & 4.27  \\
HD 11949 & 4845 & 2.85 & 1.2 & --0.09 & e & $<$10 & $<$4.6  \\
HD 12929 & 4600 & 2.70 & 1.7 & --0.13 & f & 29.9 & 4.7  \\
HD 13520 & 4043 & 1.66 & 2.1 & --0.16 & b & 157 & 4.28  \\
HD 13789 & 4740 & 4.33 & 0.8 & --0.06 & g & $<$30 & $<$5.0  \\
HD 15755 & 4611 & 2.30 & 1.2 & --0.01 & c & $<$11 & $<$4.5  \\
HD 17660 & 4713 & 4.75 & 0.8 & 0.17 & c & 31 & 5.1  \\
HD 18449 & 4340 & 2.37 & 2.0 & --0.19 & a & 34 & 4.11 \\
HD 19270 & 4723 & 2.40 & 1.5 & 0.15 & c & $<$22 & $<$5.2  \\
HD 20644 & 4100 & 1.65 & 3.0 & --0.44 & f & 93 & 4.01  \\
HD 21017 & 4443 & 2.74 & 1.9 & 0.12 & b & 40.9 & 4.67  \\
HD 26162 & 4640 & 2.87 & 2.2 & --0.02 & a & $<$9 & $<$4.4  \\
HD 26546 & 4743 & 2.25 & 1.3 & --0.01 & c & $<$20 & $<$5.0  \\
HD 27382 & 4480 & 2.67 & 2.0 & --0.37 & a & 40 & 4.47 \\
HD 29139 & 3870 & 1.66 & 2.1 & --0.04 & b & 192 & 4.23 \\
HD 29697 & 4440 & 4.19 & 0.5 & 0.18 & h & 68 & 4.8 \\
HD 30454 & 4540 & 2.60 & 1.9 & --0.31 & a & 23.6 & 4.38  \\
HD 30504 & 4056 & 1.79 & 2.1 & --0.33 & b & 136 & 4.16  \\
HD 30834 & 4219 & 1.59 & 2.3 & --0.24 & b & 134 & 4.56  \\
HD 31539 & 4210 & 2.21 & 2.1 & --0.32 & a & 57.7 & 4.06  \\
HD 32147 & 4641 & 4.60 & 1.1 & 0.26 & i & 16.8 & 4.67  \\
HD 33554 & 3970 & 1.71 & 2.0 & --0.11 & a & 197 & 4.36  \\
HD 34043 & 4120 & 2.00 & 2.2 & --0.04 & a & 48.1 & 3.95 \\
HD 34334 & 4180 & 2.12 & 1.9 & --0.46 & a & 80 & 4.0  \\
HD 35620 & 4198 & 1.92 & 2.4 & 0.15 & b & 98.7 & 4.68  \\
HD 36003 & 4345 & 4.59 & 0.5 & --0.15 & b & 16.5 & 4.16  \\
HD 37984 & 4484 & 2.21 & 2.0 & --0.41 & b & 19.8 & 4.16  \\
HD 39019 & 4770 & 2.82 & 2.1 & --0.08 & a & $<$14 & $<$4.8  \\
HD 39118 & 4550 & 1.52 & 2.2 & --0.10 & f & $<$9 & $<$4.25  \\
HD 39715 & 4798 & 4.75 & 0.9 & --0.04 & j & $<$16 & $<$4.8  \\
HD 40657 & 4300 & 1.83 & 2.3 & --0.57 & b & 29.1 & 3.79  \\
HD 43039 & 4690 & 2.81 & 2.0 & --0.33 & a & $<$5 & $<$4.2  \\
HD 47174 & 4410 & 2.30 & 2.2 & --0.10 & a & 30 & 4.24  \\
HD 47752 & 4613 & 4.60 & 0.8 & --0.05 & c & 30.2 & 4.78  \\
HD 49293 & 4620 & 2.59 & 2.3 & --0.12 & a & $<$12 & $<$4.4 \\
HD 52556 & 4700 & 2.65 & 2.3 & --0.08 & f & $<$18 & \nodata \\
HD 52960 & 4150 & 1.80 & 2.0 & --0.08 & f & 64.2 & 4.18  \\
HD 54079 & 4450 & 2.10 & 1.8 & --0.45 & f & 14.1 & 4.08  \\
HD 54716 & 4020 & 1.88 & 2.2 & --0.26 & a & 150 & 4.26  \\
HD 54719 & 4437 & 2.17 & 2.0 & 0.17 & b & 32 & 4.64  \\
HD 58207 & 4825 & 2.57 & 2.1 & --0.11 & b & $<$12 & $<$4.9  \\
HD 58972 & 4000 & 1.82 & 2.2 & --0.37 & a & 116 & 3.99  \\
HD 60522 & 3846 & 1.69 & 2.5 & 0.04 & b & 250 & 4.43  \\
HD 63752 & 4075 & 1.00 & 2.4 & --0.38 & f & 48.8 & 3.63  \\
HD 65277 & 4741 & 4.76 & 0.8 & --0.16 & j & $<$3 & $<$4.4  \\
HD 65953 & 4014 & 1.81 & 2.2 & --0.21 & b & 128 & 4.08  \\
HD 66141 & 4312 & 2.11 & 1.9 & --0.36 & b & 39.3 & 4.08 \\
HD 87883 & 4772 & 4.44 & 0.8 & 0.09 & k & $<$15 & $<$4.8  \\
HD 88320 & 3960 & 4.45 & 0.8 & --0.05 & b & 37.8 & 4.27  \\
HD 209747 & 4070 & 1.84 & 2.2 & 0.02 & a & 149 & 4.45  \\
HD 209945 & 3930 & 1.62 & 2.5 & --0.14 & a & 188 & 4.22 \\
HD 210354 & 4790 & 2.81 & 1.9 & --0.24 & a & $<$15 & $<$4.8  \\
HD 210762 & 4185 & 1.65 & 2.5 & --0.03 & f & 37.5 & 3.99  \\
HD 211075 & 4305 & 1.76 & 1.8 & --0.33 & l & 39.4 & 4.08  \\
HD 213119 & 4090 & 1.65 & 2.5 & --0.50 & f & 188 & 4.34 \\
HD 214868 & 4440 & 2.32 & 2.4 & --0.25 & a & 26.7 & 4.32  \\
HD 214995 & 4560 & 2.67 & 2.2 & --0.09 & a & $<$9 & $<$4.4  \\
HD 215182 & 5080 & 2.48 & 2.3 & --0.22 & a & $<$5 & $<$5.0  \\
HD 215665 & 4800 & 3.30 & 3.0 & --0.10 & a & $<$6 & $<$4.5  \\
HD 216174 & 4385 & 1.87 & 2.0 & --0.50 & b & 17 & 3.84  \\
HD 216646 & 4520 & 2.63 & 2.3 & --0.02 & a & 24.2 & 4.54  \\
HD 216946 & 4000 & 0.50 & 2.4 & --0.07 & m & 225 & 5.94  \\
HD 218031 & 4743 & 2.46 & 2.0 & --0.16 & b & $<$15 & $<$4.8 \\
HD 219134 & 4715 & 4.57 & 0.7 & 0.06 & b & 22 & 4.68 \\
HD 220009 & 4314 & 1.81 & 2.1 & --0.71 & b & $<$35.9 & $<$3.9 \\
HD 222107 & 4600 & 3.11 & 2.0 & --0.56 & a & 10 & 4.02  \\
HD 223559 & 3950 & 1.67 & 2.2 & --0.23 & a & 152 & 4.11 \\
HD 223807 & 4440 & 2.56 & 2.2 & --0.10 & a & $<$12 & $<$4.2  \\
HD 225212 & 4250 & 0.75 & 4.5 & --0.20 & h & 118 & 4.48  \\
HDE 233517 & 4475 & 2.25 & 1.9 & --0.37 & n & $<$35 & $<$4.5 
\enddata
\tablerefs{
(a) McWilliam 1990;
(b) Prugniel et al. 2011;
(c) Mishenina et al. 2006, 2008;
(d) Lyubimkov 2010;
(e) Takeda et al. 2008
(f) Hekker et al. 2007;
(g) Sousa et al. 2008
(h) Luck \& Bond 1980, Luck \& Heiter 2006;
(i) Allende Prieto et al. 2004;
(j) Valenti \& Fischer 2005;
(k) Kotoneva 2006;
(l) Melendez et al. 2008;
(m) Ram\'{i}rez et al. 2000;
(n) Balachandran et al. 2000.
}
\end{deluxetable}

\clearpage

\begin{deluxetable}{lcc}
\tablecaption{Sensitivity to Stellar Parameters\label{table_uncert}}
\tablewidth{0pt}
\tablehead{
  & \colhead {log $\epsilon$($F$)} & \colhead {log $\epsilon$($F$)} \\
\colhead {Parameter} & \colhead{Dwarfs} & \colhead{Giants} 
}
\startdata
$\Delta$$T_{eff}$ = +100 K  & +0.18  & +0.25  \\
$\Delta$Log $g$ = +0.3        & 0.00 & --0.02   \\ 
$\Delta$$\xi$ = +0.5 km s$^{-1}$ & 0.00 & --0.01   \\
$\Delta$[Fe/H] = +0.1  & +0.05  & +0.05 
\enddata
\end{deluxetable}

\end{document}